\documentclass[useAMS,usenatbib]{mn2e}
\usepackage{aas_macros}
\usepackage{layout}
\usepackage[dvips]{graphicx}
\title[Dark Matter Haloes in the Cosmic Web]{Properties of Dark Matter Haloes in Clusters, Filaments, Sheets and Voids}
\author[O. Hahn et al.]{ Oliver Hahn$^{1}$,
Cristiano Porciani$^{1}$, C. Marcella Carollo$^{1}$ and Avishai Dekel$^{2}$\\
$^{1}$ETH Z\"urich, 8093 Z\"urich, Switzerland \\
$^{2}$Racah Institute of Physics, The Hebrew University, Jerusalem, Israel}
\begin{document}

\date{MNRAS in press.}
\pagerange{\pageref{firstpage}--\pageref{lastpage}} \pubyear{2006}
\maketitle

\label{firstpage}

\begin{abstract}
Using a series of high-resolution N-body simulations of the concordance cosmology 
we investigate how the formation histories, shapes
and angular momenta of dark-matter haloes depend on environment.
We first present a classification scheme that allows to distinguish between haloes in 
clusters, filaments, sheets and voids in the large-scale distribution of matter. 
This method (which goes beyond a simple measure of the local density)
is based on a local-stability criterion for the orbits of test particles and
closely relates to the Zel'dovich approximation.
Applying this scheme to our simulations we then find that:
{\it i)}
Mass assembly histories and formation redshifts strongly depend on environment for
haloes of mass $M<M_*$
(haloes of a given mass tend to be older in clusters and younger in voids)
and are independent of it for larger masses ($M_*$ here indicates the
typical mass scale which is entering the non-linear regime of perturbation
growth);
{\it ii)}
Low-mass haloes in 
clusters are generally less spherical and more oblate than in other regions;
{\it iii)}
Low-mass haloes in clusters have a higher median spin than in filaments
and present a more prominent fraction of rapidly spinning objects;
we identify recent major mergers as a likely source of this effect.
For all these relations, we 
provide accurate functional fits as a function of halo mass and environment.
We also look for correlations between halo-spin directions and the large-scale structures:
the strongest effect is seen in sheets where halo spins tend to lie within the plane of symmetry
of the mass distribution.
Finally, we measure the spatial auto-correlation of spin directions and the cross-correlation between 
the directions of intrinsic and orbital angular momenta of neighbouring haloes.
While the first quantity is always very small, we find that spin-orbit correlations are rather
strong especially for low-mass haloes in clusters and high-mass haloes in filaments.
\end{abstract}


\section{Introduction}

Numerical simulations and analytical work have shown that the gravitational amplification of small density fluctuations 
leads to a wealth of structures resembling the observed large-scale distribution of galaxies.
The resulting mass density distribution can be thought of as a ``cosmic web'' \citep{Bond96}
characterised by the presence of structures with different dimensionality.
Most of the volume resides in low-density regions (voids) which are surrounded by thin denser sheets of matter.
A network of filaments of different sizes and density contrasts departs from the sheets and visually dominates
the mass distribution.
Dense clumps of matter lie at the intersections of filaments.
From the dynamical point of view, matter tends to flow out of the voids, transit through the sheets and 
finally accrete onto the largest clumps through the filaments.

In a Universe dominated by cold dark matter (CDM), this description applies only after coarse-graining
the density distribution on scales of a few Mpc.
On smaller scales, the power in the primordial spectrum ends up producing a hierarchical distribution 
of (virialised) dark-matter haloes whose positions trace the large-scale structure
described above. According to the current cosmological paradigm, galaxies form within these haloes.

Astronomical observations show that galaxy properties in the local Universe vary systematically with
environment \citep[e.g.][]{Dressler80, Kauffmann04, Blanton05}. 
As a fundamental step towards understanding galaxy formation it is thus important
to establish how the properties of dark-matter haloes depend on the environment in which they reside.
A first attempt in this direction was made by \cite{Lemson1999} who found that mass is the only halo property that correlates with environment at variance with concentration, spin, shape and formation epoch.
Using marked statistics, \cite{Sheth2004} found evidence that haloes of a given mass form earlier in dense regions.
Higher resolution simulations confirmed this finding and helped to better quantify it as a function of halo mass
and redshift \citep{Gao2005,Croton06,Harker2006,Reed06,Maulbetsch06}.
At the same time it has become clear that also other halo properties as concentration and spin correlate
with local environment \citep{AvilaReese2005,Wechsler05,Bett06,Maccio06,Wetzel06}.

Although the large-scale structure of matter is prominently reflected in the halo distribution, no efficient automated method has been proposed to associate a given halo to the dynamical structure it belongs to. Most of the environmental studies mentioned above use the local mass density within a few Mpc as a proxy for environment. In this paper we follow a novel approach and associate dark-matter haloes to structures with different dynamics. 
{\it Voids}, {\it sheets}, {\it filaments} and {\it clusters} are distinguished based on a stability criterion for the orbit of test particles which is inspired by the Zel'dovich approximation \citep{Zeldovich70}. Our method is accurate, fast, efficient and contains only one free parameter which fixes the spatial resolution with which the density field has to be smoothed (as in the evaluation of the density). We show that any classification based on local density is degenerate with respect to ours which we regard as more fundamental. We find that
{\it all} halo properties at zero redshift show some dependence on the dynamical environment in which they reside. We accurately quantify this dependence and show that halo properties smoothly change when one moves from voids to sheets, then to filaments and finally to clusters. Redshift evolution of these trends will be investigated in future work.

This paper is organised as follows. 
In Section \ref{sec:Nbody}, we briefly describe the N-body simulations we use and how we compute a number of halo properties. The method for the identification of the halo environment is presented in Section \ref{sec:Classification} together with a number of tests that show how well the method performs. Our main results on the environmental dependence of the halo properties are given in Section \ref{main}. Finally, Section \ref{conclusions} summarises the main conclusions that we draw from our work.

\section{N-Body Simulations}
\label{sec:Nbody}
We used the tree-PM code GADGET-2 \citep{Springel2005} to follow the formation and the evolution of the large-scale structure in a flat $\Lambda$CDM cosmology. We have assumed the matter density parameter $\Omega_{\rm m}=0.25$, with a baryonic contribution
$\Omega_{\rm b}=0.045$, and the present-day value for the Hubble constant $H_0=100 h$ km s$^{-1}$ Mpc$^{-1}$, with $h=0.73$.
In particular, we performed three N-body simulations, each containing $512^3$ dark matter particles in periodic boxes of size $L_1=45\, h^{-1}$ Mpc, $L_2=90\, h^{-1}$ Mpc and $L_3=180\,h^{-1}$ Mpc. The corresponding particle masses are $4.7\times 10^7\,h^{-1}M_\odot$, $3.8\times 10^8\,h^{-1}M_\odot$ and $3.0\times 10^9\,h^{-1}M_\odot$, respectively.
The simulations follow the evolution of Gaussian density fluctuations characterised by a scale-free initial power spectrum with spectral index $n=1$ and normalisation $\sigma_8=0.9$ (with $\sigma_8$ the rms linear density fluctuation within a sphere of $8 \,h^{-1}$ Mpc comoving radius). 
The initial conditions were generated using the GRAFIC2 tool \citep{Bertschinger2001} for the redshift $z$ at which the rms density fluctuation on the smallest resolvable scale in each box equals 0.1. This corresponds to $z\simeq 79, 65$ and 52 for $L_1, L_2$ and $L_3$ respectively. 
Particle positions and velocities were saved for 30 time-steps logarithmically spaced in expansion parameter $a$ between $z=10$ and $z=0$.

\subsection{Halo identification and properties}
Virialised dark-matter haloes were identified using the standard friends-of-friends (FOF) algorithm with a linking length equal to $0.2$ times the mean inter-particle distance. 
We only considered haloes containing at least 300 particles, since virtually all of the halo properties we investigated show strong numerical artefacts when measured for less well resolved haloes. We found 13353, 16296 and 21041 of such haloes for $L_1$, $L_2$ and $L_3$, respectively. The most massive groups in the three simulations contain nearly $10^{6-7}$ particles and have masses $4.3\times 10^{14} h^{-1} M_\odot$, $7.6\times 10^{14} h^{-1} M_\odot$ and $2.2\times 10^{15} h^{-1} M_\odot$. Our catalog therefore spans five orders of magnitude in halo mass with high resolution haloes, ranging from the size of dwarf galaxies to massive clusters.

We characterised the mass assembly and merging history of the halos as follows. For each halo at redshift $z$, we identified a progenitor at $z_{\rm p}>z$ by intersecting the sets of their particles. The main progenitor was then chosen to be the most massive halo at each redshift that contributes at least 50 per cent of its particles to the final halo. We then defined the formation redshift $z_{\rm form}$ as the epoch at which a main progenitor which has at least half of the final mass first appears in the simulation and interpolated linearly between simulation snapshots in $\log z$ to find the point where exactly half of the mass is accumulated. 

%%%

\subsubsection{Halo Shapes}
\label{sec:HaloShapes}
In order to quantify the shape of FOF haloes, we determined their moment of inertia tensor, defined as
\begin{equation}
I_{jk} \equiv m\sum_{i}\left(r_i^2\delta_{jk}-x_{i,j}x_{i,k}\right),
\end{equation}
where $m$ is the particle mass, $r_i\equiv(x_{i,1},x_{i,2},x_{i,3})$ is the distance of the $i$-th particle from the centre of mass of the halo and $\delta_{jk}$ denotes the Kronecker symbol. The eigenvectors of ${\bf I}$ are related to the lengths of the principal axes of inertia $l_1\geq l_2\geq l_3$ \citep[e.g.][]{Bett06}. We used the following dimensionless quantities
\begin{equation}
S = \frac{l_3}{l_1}\qquad \textrm{and}\qquad
T  = \frac{l_1^2-l_2^2}{l_1^2-l_3^2}
\end{equation}
to measure sphericity and triaxiality of the haloes \citep[e.g.][]{Franx1991, Warren1992}. A spherical halo has $S=1$, a needle $S=0$, a prolate halo $T=1$ and an oblate one $T=0$.

%%%

\subsubsection{Halo Spin Parameter}
\label{sec:SpinParam}
The spin parameter of a halo is a dimensionless quantity introduced by \cite{Peebles1969} that indicates the amount of ordered rotation compared to the internal random motions. For a halo of mass $M$ and angular momentum ${\mathbf J}$ it is defined as
\begin{eqnarray}
\lambda & = & \frac{\left|\mathbf{J}\right|\,\left|E\right|^{1/2}}{GM^{5/2}},
\end{eqnarray}
where the total energy $E=T+U$ with $T$ the kinetic energy of the halo after subtracting its bulk motion and $U$ the potential energy of the halo produced by its own mass distribution.
Determining the potential energy of massive haloes is computationally expensive, so \cite{Bullock2001} introduced the alternative spin parameter
\begin{equation}
\lambda^{\prime} \equiv \frac{\left|\mathbf{J}_{\rm vir}\right|}{\sqrt{2}M_{\rm vir}V_{\rm vir} R_{\rm vir}}.
\end{equation}
Here all quantities with the subscript ``vir'' (angular momentum, mass and circular velocity) are computed within a sphere of radius $R_{\rm vir}$ which approximates the virial radius of the halo. As this quantity is not well defined for FOF groups, we took $R_{\rm vir}$ to be a fraction $\alpha$ of the maximum distance between a halo particle and the centre of mass.
To accommodate possible fuzzy boundaries of the haloes, we chose a value of $\alpha=0.95$. We verified that the particular choice of $\alpha$ does not have an impact on the distribution of $\lambda^{\prime}$ 
and remains unchanged even when going as low as $\alpha=0.1$ \citep[see also][]{Bullock2001}.
Under the assumption that the halo is in dynamical equilibrium, $V_{\rm vir}^{2}=GM_{\rm vir}/R_{\rm vir}$, 
the spin parameter can be rewritten as
\begin{equation}
\lambda^{\prime} = \frac{\left|\mathbf{J}_{\rm vir}\right|} {\sqrt{2GR_{\rm vir}} M_{\rm vir}^{3/2}}.
\end{equation}
We found a spurious increase in $\lambda^{\prime}$ for haloes consisting of less than 250-300 particles. This numerical effect occurred for all of our three simulated boxes. The median spin $\lambda^{\prime}_{\rm med}$ is roughly 10 per cent higher for haloes with only 100 particles than for haloes consisting of more than 300 particles.

%%%

\section{Orbit stability  and environment}
\label{sec:Classification}
\subsection{Basic theory}
\label{sec:ClassBasic}
We use a simple stability criterion from the theory of dynamical systems to 
distinguish between haloes residing in clusters, filaments, sheets or voids.
Consider a test particle moving in the peculiar gravitational potential, $\phi$, generated by a cosmological matter distribution frozen in time (e.g. no Hubble drag). The equation of motion in comoving coordinates for this test particle is $\ddot{\mathbf{x}}  =  -\mathbf{\nabla}\phi$, where the dot represents derivatives with respect to a fictitious time. Assuming that at the centre of mass of each halo $\bar{\mathbf{x}}_i$ the gravitational potential has a local extremum (i.e. $\mathbf{\nabla}\phi(\bar{\mathbf{x}}_i)=0$), the fixed points of the test particle equation of motion are exactly at the points $\bar{\mathbf{x}}_i$. We can thus linearise the equation of motion at the points $\bar{\mathbf{x}}_i$ and find the linear system
\begin{eqnarray}
\ddot{x}_i & = & -T_{ij}(\bar\mathbf{x}_k)\,(x_j-\bar{x}_{k,j})\,,
\end{eqnarray}
where the tidal field $T_{ij}$ is given by the Hessian of the gravitational potential
\begin{eqnarray}
T_{ij}& \equiv & \partial_i\partial_j\,\phi\;. 
\end{eqnarray}
Thus the linear dynamics near local extrema of the gravitational potential is fully governed by the three 
(purely real, as $T_{ij}$ is symmetric) eigenvalues of the tidal field tensor. We use the number of positive eigenvalues of $T_{ij}$ to classify the four possible environments a halo may reside in. Note that the number of positive eigenvalues is equivalent to the dimension of the stable manifold at the fixed points. In analogy with Zel'dovich theory
\citep{Zeldovich70}, we define as
\begin{enumerate}
  \item {\it voids} the region of space where $T_{ij}$ has no positive eigenvalues (unstable orbits);
  \item {\it sheets} the set of points with one positive and two negative eigenvalues (1-dimensional stable manifold);
  \item {\it filaments} the sites with two positive and one negative eigenvalue (2-dimensional stable manifold);
  \item {\it clusters} the zones with three positive eigenvalues (attractive fixed points).
\end{enumerate}
Dropping the assumption of local extrema of the gravitational potential at the centres of mass of the haloes introduces a constant acceleration term to the linearised equations of motion. This zeroth-order effect can be dispersed of by changing to free-falling coordinates. The deformation behaviour introduced by the first-order term, however, remains unchanged.
\begin{figure}[h!]
  \begin{center}
    \includegraphics[width=0.45\textwidth]{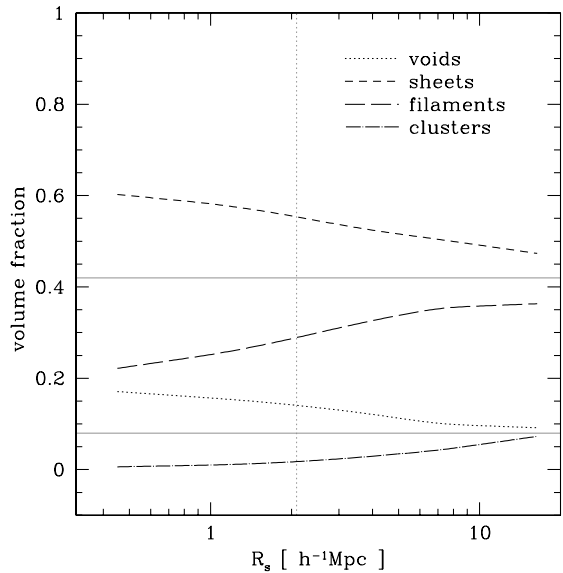}
  \end{center}
  \caption{ \label{fig:StructureFraction}The volume fraction being classified as clusters, filaments, sheets or voids for our $180\,h^{-1}$ Mpc box as a function of the smoothing scale $R_{\rm s}$. The vertical dotted line at $R_{\rm s}=2.1 h^{-1}$ Mpc indicates the smoothing scale adopted in this paper. The solid grey lines indicate the predicted volume fractions for a Gaussian random field \citep{Doro70}. For very large $R_{\rm s}$, the non-Gaussian density field of the simulations asymptotes to the predicted fractions of 42 per cent for sheets and filaments and 8 per cent for voids and clusters. Volume fractions are evaluated on a $128^3$ Cartesian subgrid.}
\end{figure}
\subsection{Implementation}
In order to determine the eigenspace structure of the tidal field tensor, we first compute the peculiar gravitational potential $\phi$ from the matter density distribution via Poisson's equation
\begin{eqnarray}
\nabla^2\phi & = & {4\pi G \,\bar{\rho}}\,\delta,
\end{eqnarray}
where $\bar{\rho}$ and $\delta$ respectively denote the mean mass density of the universe and the overdensity field. For our N-body simulations, we solve Poisson's equation using a fast Fourier transform on a grid of twice the particle resolution ($1024^3$ grid cells). The density field $\delta$ is obtained by using Cloud-In-Cell interpolation of the particles onto the grid and then smoothed using a Gaussian kernel
$K_{R_{\rm s}}$. In this case, the smoothing length, $R_{\rm s}$,  and the mean mass contained in the filter, $M_{\rm s}$, follow the relation 
\begin{eqnarray}
R_{\rm s} & = & \frac{1}{\sqrt{2\pi}}\left(\frac{M_{\rm s}}{\bar{\rho}}\right)^{1/3}\;.
\end{eqnarray}
To solve Poisson's equation on the grid, we apply the Green's function $G^{(2)}$ of the symmetric 5-point finite difference operator that we later use to compute the tidal tensor.
Altogether, we hence find the solution for the smoothed gravitational potential through the double convolution
\begin{eqnarray}
\phi_{R_{\rm s}} & = & \delta\star K_{R_{\rm s}}\star G^{(2)}\;.
\end{eqnarray}
We then apply the second derivative operator to $\phi_{R_{\rm s}}$ and get the diagonal components of the tidal tensor. For the off-trace components, we apply twice the symmetric first derivative operator in the corresponding coordinates. Although the second derivative operator cannot be produced from applying twice a symmetric first derivative operator, following this scheme ensures that the trace coincides with the smoothed overdensity to machine accuracy, while the off-trace components are indeed symmetric and are not suffering from a spurious self-potential. Finally, we compute the eigenspace structure of the tensor at each halo's centre of mass.

\begin{figure*}
  \vspace*{-15pt}
  \begin{center}
    \includegraphics[width=0.8\textwidth]{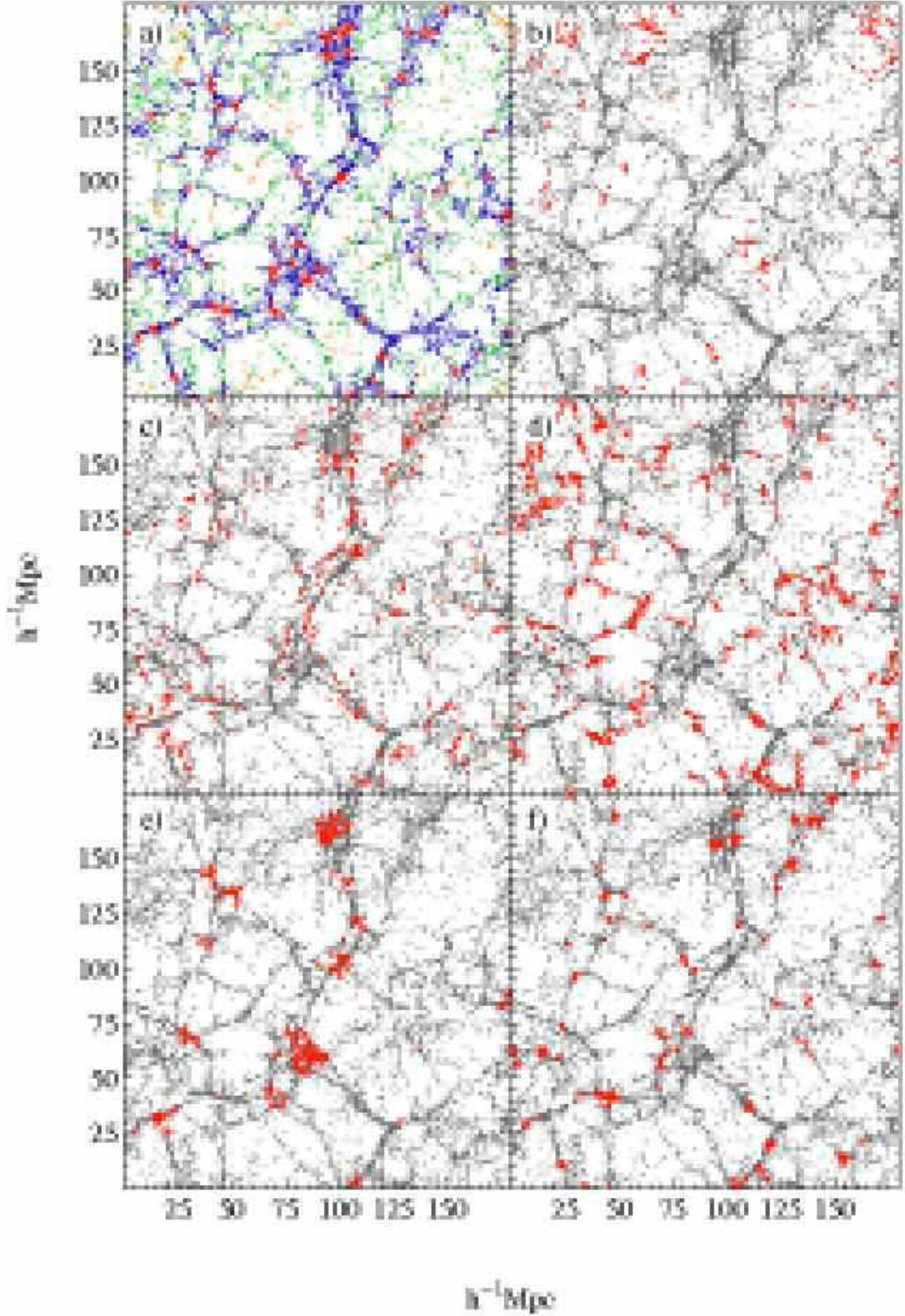}
  \end{center}
  \vspace*{-20pt}
  \caption{ \label{fig:Transitionobjects} Classification of halo environments in a slice of 10$h^{-1}$ Mpc thickness
for the 180$h^{-1}$Mpc box. 
Panel {\it a)} shows the classification scheme for a smoothing scale of $R_{\rm s}=2.1 h^{-1}$Mpc with the following colour coding: clusters (red), filaments (blue), sheets (green) and voids (orange). Panels {\it b)} to {\it f)} show in red those haloes that change classification in a specific way when the smoothing scale is increased to $R_{\rm s}=4.5 h^{-1}$Mpc, all other haloes are represented in grey. Panel {\it b)} represents sheets at smaller smoothing $R_{\rm s}=2.1 h^{-1}$Mpc that become voids at larger smoothing $R_{\rm s}=4.5 h^{-1}$Mpc. Panel {\it c)} shows sheets that become filaments, panel {\it d)} filaments that become sheets, panel {\it e)} filaments that become clusters and panel {\it f)} shows clusters that become filaments. To achieve higher spatial resolution in the visual representations, all haloes down to 10 particles are shown in the panels above.}
\end{figure*}

%%%

\subsection{Optimisation}
\label{sec:Optimisation}
\begin{figure}
  \begin{center}
    \includegraphics[width=0.45\textwidth]{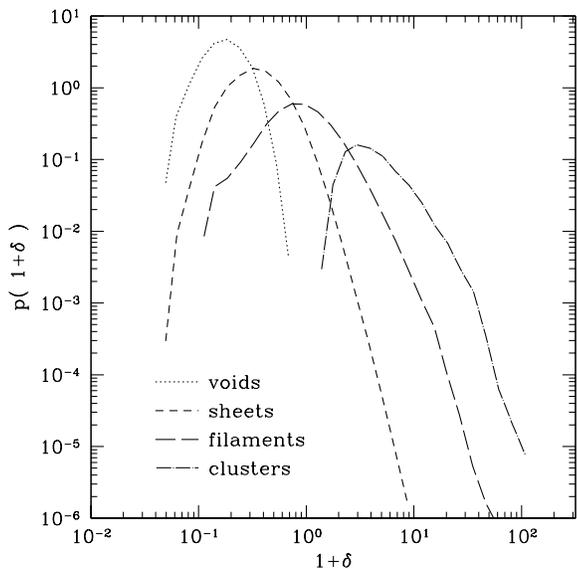}
  \end{center}
 \caption{ \label{fig:DensityStat}Volume weighted probability distribution of the local density for clusters, filaments, sheets and voids. Statistics have been obtained combining all three simulation volumes. Statistics weighted by halo abundance shifts the distributions to over-densities roughly a factor of 2 higher. Note that the stability criterion naturally finds ``clusters'' in the highest density regions and ``voids'' in the lowest and thus disambiguates any definition of environment that is solely based on density measures.} 
\end{figure}

Our criterion for determining the halo environment contains one free parameter, namely the smoothing radius of the Gaussian kernel, $R_{\rm s}$.
This corresponds to the typical length-scale over which we determine the dynamical stability of the orbits. 
The particular choice of $R_{\rm s}$ directly affects the local eigenstructure and thus changes the classification of environment. Smoothing on the scale of single haloes picks out each single halo as a stable cluster in the sense of the definition. 
In Figure \ref{fig:StructureFraction} we show how the choice of $R_{\rm s}$ affects the fraction of the simulated volume classified in the four categories. These fractions continuously vary with $R_{\rm s}$, which implies that some haloes change their classification. 
For $R_{\rm s}\gg 10 h^{-1}$ Mpc, the density field becomes approximately Gaussian, and we observe convergence to the theoretical volume fractions of 42 per cent for sheets and filaments, and 8 per cent for voids and clusters 
(Doroshkevich 1970, see also Shen et al. 2006).
\nocite{Doro70,Shen06} \\
To illustrate the transition of haloes between environment classes, in Figure \ref{fig:Transitionobjects} we use a snapshot at $z=0$ to highlight the haloes that are assigned to different environments when the smoothing scale is changed from $2.1 \, h^{-1}$ Mpc to $4.5\, h^{-1}$ Mpc (corresponding to a change by a factor of 10 in $M_{\rm s}$). Basically, increasing the smoothing scale:
{\it i)} increases the number of haloes in voids at the expenses of the surrounding sheets (panel b);
{\it ii)} moves the thin filaments surrounding a thicker one from the sheet environment to the filament one (panel c);
{\it iii)} moves the thin filaments surrounding a void from the filament environment to the sheet one (panel d);
{\it iv)} increases the size of massive clusters located at the intersection of filaments at the expenses of the ending points of filaments themselves (panel e); 
{\it v)} moves the densest clumps located along filaments from the cluster environment to the filament one (panel f). 
Table \ref{matrix} lists the fraction of the total number of haloes that are assigned to the 16 possible classifications with the two smoothing scales. The halos that contribute to the off-diagonal elements of this ``transition matrix'' typically live in regions where the tidal field has one nearly vanishing eigenvalue. In these transition regions, a modification of $R_{\rm s}$ can easily change the sign of this eigenvalue and thus the association of the corresponding halo to its environment. This results from using sharp boundaries (positive vs negative eigenvalues) to classify the different environments. Note that only a negligible fraction of the haloes inverts the sign of more than one eigenvalue of the tidal field when the smoothing scale is changed, indicating that our classification is indeed physical.
Based on Figure \ref{fig:Transitionobjects}, we conclude that
the combined use of two (or more) smoothing scales can be used to classify a larger variety of environments with respect to the basic four that can be found with a fixed resolution, and in particular to identify boundary regions that bridge between the basic four types. We will explore this potentiality of the orbit-stability method in future works. For simplicity, in this paper we only consider a single smoothing scale, $R_{\rm s}=2.1 \, h^{-1}$ Mpc (corresponding to $M_{\rm s}\approx10^{13} h^{-1} M_\odot$) which provides startling agreement between the outcome of the orbit-stability criterion and a visual classification of the large-scale structure. The resulting classification of halo environments is highlighted in the top-left panel of Figure \ref{fig:Transitionobjects} using different colours. 
For $R_{\rm s}=2.1 \, h^{-1}$ Mpc, the volume fractions occupied by voids, sheets, filaments and clusters are, respectively, 13.5\%, 53.6\%, 31.2\% and 1.7\%. This suggests that we identify as voids just the inner parts of the most under-dense regions (see also Figure \ref{fig:DensityStat}) and consider as sheets the volume-filling regions around them.
At the same time, our clusters always contain haloes with a virial mass $M_{\rm vir}>10^{13} h^{-1} M_\odot$ and, in some cases, haloes with $M_{\rm vir}>10^{14} h^{-1} M_\odot$ and radius $R_{\rm vir}> 1\, h^{-1}$ Mpc (which are usually tagged as clusters). These haloes typically constitute the central parts of what we identify as clusters. By definition, our ``cluster environment'' extends to distances which are significantly larger than $R_{\rm vir}$ and also includes all the smaller haloes that are infalling onto or orbiting around the central one. For the value of $R_{\rm s}$ adopted in this paper, we find that ``our'' clusters have typical diameters of a few Mpc. We have tested that all our findings do not depend on the precise choice of $R_{\rm s}$.
\begin{table}
\caption{\label{matrix}Transition matrix for halo classification between smoothing at
$M_{\rm s}=10^{13}M_{\odot}$, indicated by ``(S)'', and
$M_{\rm s}=10^{14}M_{\odot}$, indicated by ``(L)''. Matrix entries are given in per cent
of the total number of haloes. Non-diagonal elements represent
haloes that change classification.}
\begin{tabular}{@{}rcccc@{}}
\hline
{} & void (L) & sheet (L) & filament (L) & cluster (L) \\
\hline
void (S) & 0.06 & $<$0.01 & 0 & 0 \\
sheet (S) & 0.63 & 10.4 & 2.9 & 0.01 \\
filament (S) & 0.41 & 15.1 & 46.5 & 7.3 \\
cluster (S) & 0.02 & 1.9 & 8.7 & 5.8 \\
\end{tabular}
\end{table}

%%%

\subsection{Orbit stability vs density}

Most of the work on the environmental dependence of halo properties has hitherto considered the local density as a measure of environment \citep[e.g.][]{Lemson1999,Maccio06,Maulbetsch06}. Density corresponds to the trace of the tidal field tensor $T_{ij}$, and thus provides more limited information regarding the dynamical properties of the local flow compared to our classification, which is based on all three eigenvalues.
In Fig. \ref{fig:DensityStat} we show that local overdensity is largely degenerate relative to the four categories we derive from the eigenstructure. Density correlates with the dimension of the stable manifold, e.g.  
the median overdensity in each environment is -0.79, -0.55, 0.28 and 4.44 for voids, sheets,
filaments and clusters, respectively. 
However, it is not possible to recover, from the density field, the more detailed environmental information that we derive from the tidal field tensor. A simple environmental classification that is based on density therefore mixes our halo populations.
%%%

\section{Halo Properties and environment}
\label{main}
In this section we present a detailed study of halo properties at $z=0$ as a function of the cluster, filament, sheet and void environment determined by our orbit stability criterion.
\subsection{Mass function}
\begin{figure}
  \begin{center}
    \includegraphics[width=0.45\textwidth]{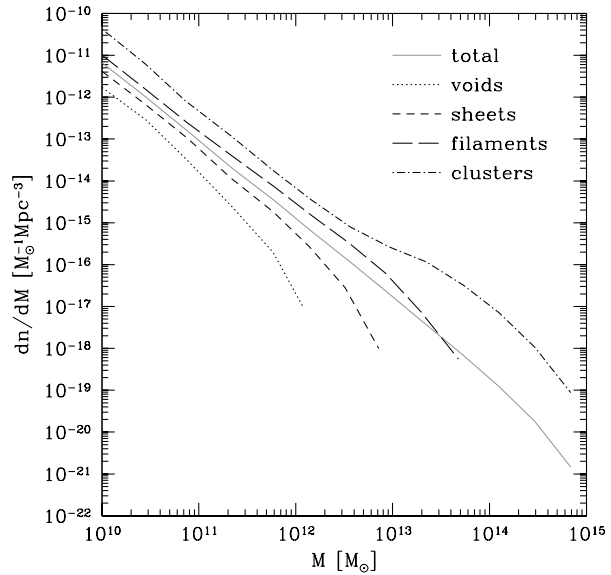}
  \end{center}
  \caption{ \label{fig:massfunction} Mass function of the haloes residing in voids, sheets, filaments and clusters. Abundances in the whole box have been rescaled by the corresponding volume fractions. The solid grey line represents the total mass function, not split into environments. Haloes from all three simulations are included. The total mass function perfectly coincides with common analytic fits \citep[e.g.][]{Jenkins01}.}
\end{figure}
Figure \ref{fig:massfunction} presents the mass functions of the haloes residing in the different environments. The low-mass end has the same slope in all environments, but the position of the high-mass cutoff is a strong function of environment. The cluster mass function is top-heavy with respect to voids, while filaments and sheets lie in between. As expected, the mean halo density is higher in clusters and lower in voids. All this is in good qualitative agreement with the conditional mass function as a function of local density derived from analytic models \citep{BCEK,Bower91}.

%%%
\begin{figure*}
  \begin{center}
    \includegraphics[width=0.45\textwidth]{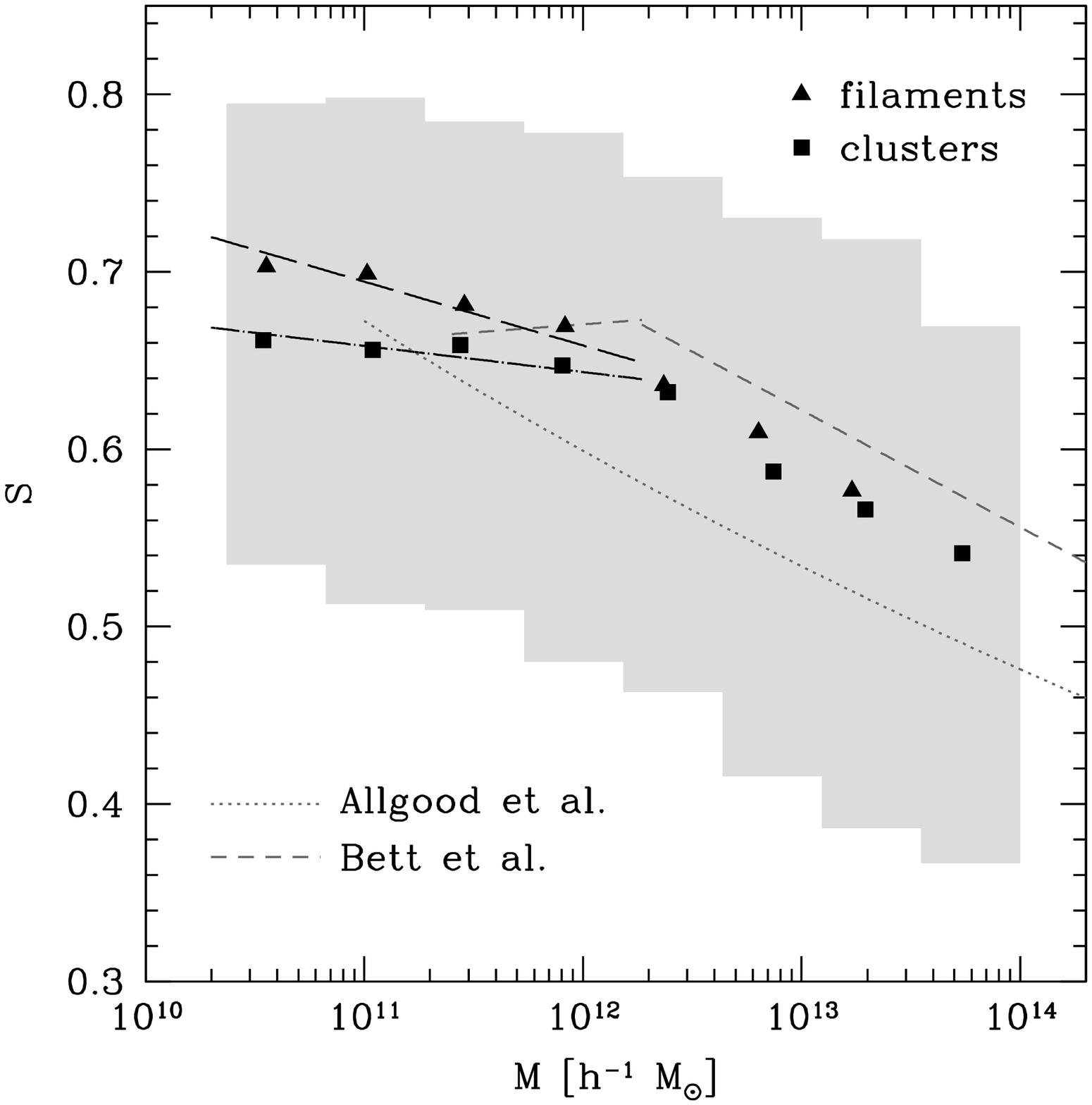}
    \includegraphics[width=0.45\textwidth]{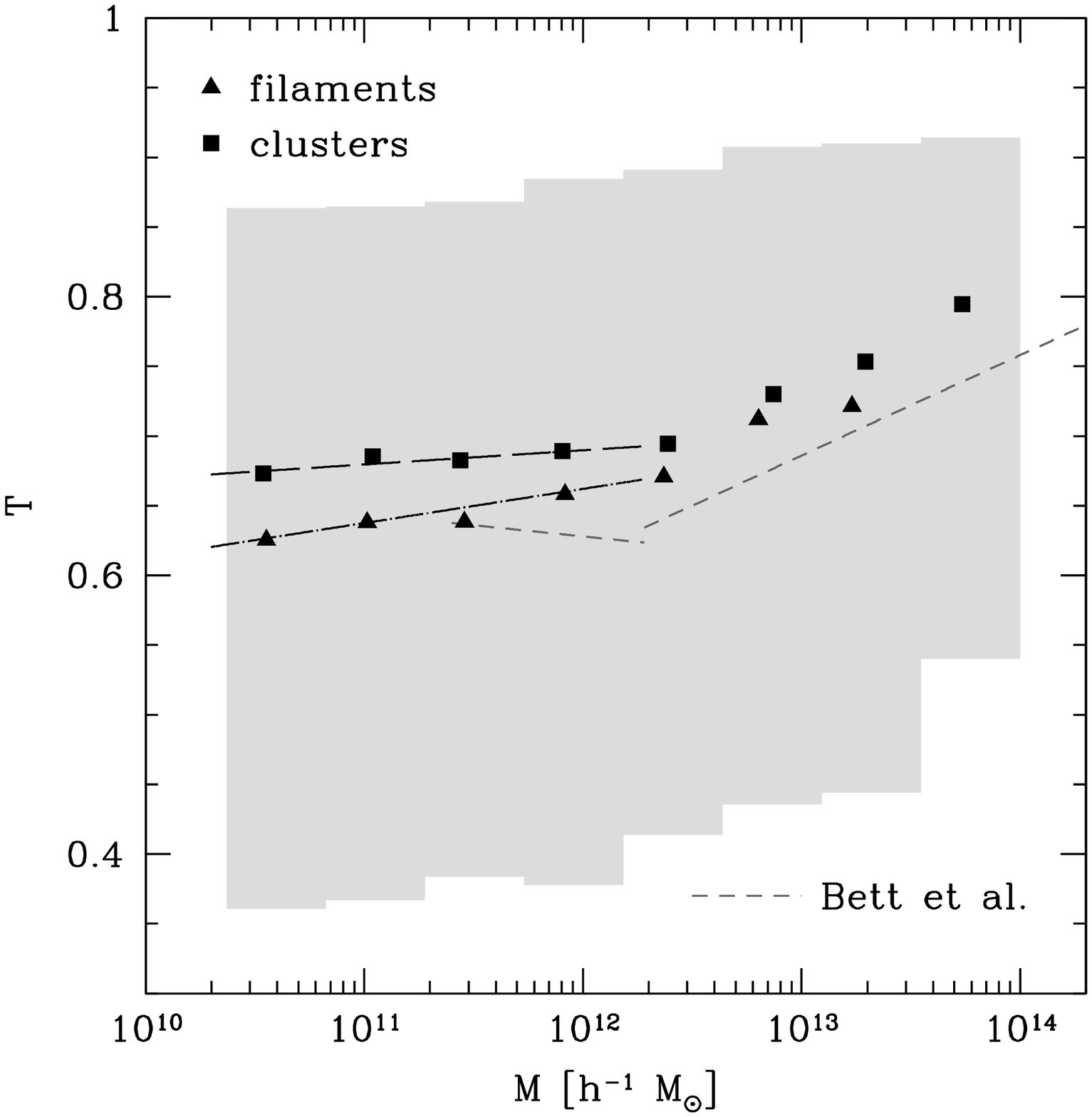}
  \end{center}
\caption{Median halo sphericity (left) and triaxiality (right) as a function of halo mass for haloes in filaments and clusters. The behaviour for haloes in sheets is almost identical to that for filaments. The shaded area indicates the central $1\sigma$ scatter in the whole sample, not split by environment. The dark grey lines indicate the fits of \citet{Allgood06} for $S$ and \citet{Bett06} for $S$ and $T$, the black lines show our fits to haloes with masses $M<2\times10^{12}h^{-1}M_{\odot}$. Parameters are given in section \ref{sec:Shapes}.}
  \label{fig:Shapes}
\end{figure*}
\subsection{Halo Shapes}
\label{sec:Shapes}
\begin{figure}
  \begin{center}
    \includegraphics[width=0.45\textwidth]{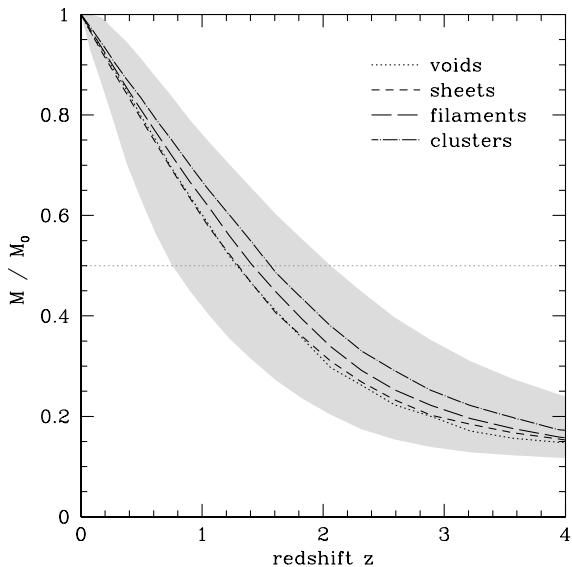}
  \end{center}
  \caption{ \label{fig:AssemblyHistory}Median mass of the main progenitor of haloes in the mass range $5\times10^{10}h^{-1}M_{\odot} < M < 5\times10^{11}h{-1}M_{\odot}$ over redshift as a fraction of the mass at $z=0$ for the four different environments. The shaded area indicates the 1$\sigma$ spread of haloes in filament environments. The spread is slightly larger for haloes in clusters. The dotted grey line indicates where $z_{\rm form}$ is measured.}
\end{figure}
 Figure \ref{fig:Shapes} shows the median of the shape parameters $S$ and $T$  for haloes in filaments and clusters as a function of their mass (the void sample contains too few haloes and the sample for sheets shows an identical behaviour to the filaments). Only when requiring that haloes in the samples contain at least 500 particles we find convergence of the median shape parameters at the lower mass end.
The overall mass dependence of $S$ and $T$ is in good agreement with previous studies \citep[e.g.][]{Allgood06,Altay06,Bett06,Maccio06}. \citet{Allgood06} fit a power law to the median $S$ as a function of halo mass, while \citet{Bett06} detect a change in slope at masses $M_c\approx2\times10^{12}h^{-1}M_\odot$. This breakpoint $M_{\rm c}$ is present also in our findings. Interestingly, it coincides with the mass above which we do not find any significant dependence of the shape parameters on environment. Our results agree very well with the measured slopes of both fitting formulas for masses $M>M_{\rm c}$. \citet{Bett06} argue that the offset of their fit with respect to \citet{Allgood06} results from different halo finding algorithms which also explains why our haloes are slightly less spherical. We do not find evidence for decreasing sphericity at the low-mass end as indicated by \citet{Bett06}, based on haloes with less than 300 particles. However, we clearly detect a decrease in slope for halo masses $M<10^{12}h^{-1} M_\odot$ with respect to the fitting formula of \citet{Allgood06}.\\
The vast dynamic range of our suite of simulations allows the unprecedented exploration of the low-mass end with high-resolution haloes (500 particles for a $2\times 10^{10} h^{-1}M_\odot$ halo).  For masses in the range $2\times10^{10}h^{-1}M_\odot<M<M_c$ we detect a clear dependence on environment. Haloes residing in clusters tend to be less spherical and more prolate almost independently of mass. In contrast, haloes in filaments tend to be slightly more oblate as one might expect from accretion of matter onto the filament. The difference between the two classes are, however, small with respect to the intrinsic scatter. For masses $M<M_c$, our results are well described by a fit of the following form:
\begin{eqnarray}
S_{\rm med} & = & s_1\,+\,\frac{s_2}{100}\,\log_{10}\frac{M}{10^{12}h^{-1}M_\odot}\\
T_{\rm med} & = & t_1\,+\,\frac{t_2}{100}\,\log_{10}\frac{M}{10^{12}h^{-1}M_\odot}.
\end{eqnarray}
\begin{eqnarray}
\left. \begin{array}{rrl}
s_1 & = & 0.66\pm0.08,\\
s_2 & = & -3.6\pm0.7;  \\
t_1 & = & 0.66\pm0.03,\\
t_2 & = & 2.4\pm0.24
\end{array} \right\} & \vspace{-6pt}\textrm{in filaments,} & \hspace{-5pt}M<M_{\rm c},\\
\left. \begin{array}{rrl}
s_1 & = & 0.64\pm0.05,\\
s_2 & = & -1.5\pm0.42;  \\
t_1 & = & 0.69\pm0.03,\\
t_2 & = & 1.0\pm0.24
\end{array} \right\} & \vspace{-6pt}\textrm{in clusters,} & \hspace{-5pt}M<M_{\rm c},
\end{eqnarray}
where $M_{\rm c}=2\times10^{12}h^{-1}M_\odot$. These values are obtained with a robust iterative least-squares fit using a bisquare estimator.

%%%

%
\subsection{Assembly history and formation redshift}
\label{sec:FormationTimes}

In Figure \ref{fig:AssemblyHistory} we show the assembly history of haloes with masses $5\times10^{10}h^{-1}M_{\odot} < M < 5\times10^{11}h^{-1}M_{\odot}$ in the different environments. In particular, we plot the median mass of the main progenitor as a function of redshift at which it is identified. The shaded area indicates the central $1\sigma$ spread for haloes in the filament environment. Although haloes tend to assemble their mass earlier in clusters and later in voids, the effect is relatively small with respect to the intrinsic scatter. This is in very good agreement with the findings of \citet{Maulbetsch06}. These authors investigated the mass assembly history splitting the halo sample by density, smoothed on $4h^{-1}$Mpc. Their high density sample ($\delta>5$) roughly corresponds to our densest clusters, while the low density sample ($\delta<0$) includes voids, sheets and the lower density filaments.
\begin{figure}
  \begin{center}
    \includegraphics[width=0.45\textwidth]{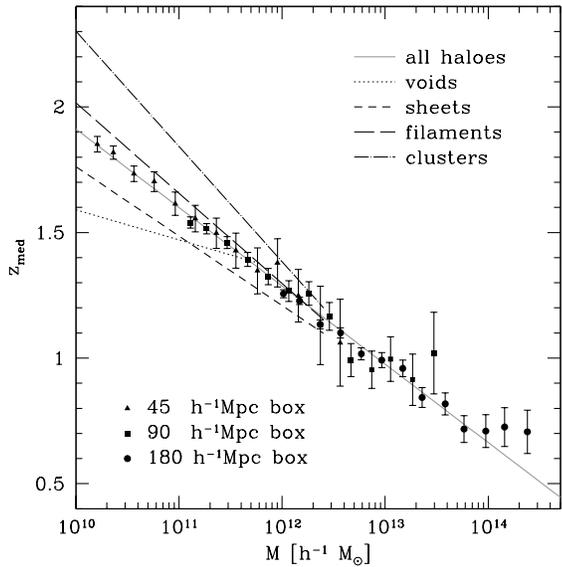}
  \end{center}
  \caption{The median formation redshift $z_{\rm med}$ for haloes from our three simulated boxes as a function of their mass. Errorbars indicate the error in the median. The grey line indicates the result of a robust fit to the displayed medians. For masses below $5\times10^{12}h^{-1}\,M_{\odot}$ we find a strong correlation with our definition of environment. The black lines indicate robust fits to the values for haloes with masses $M<5\times10^{12}h^{-1}\,M_{\odot}$ that reside in the corresponding environments. The fit parameters for all environments are given in section \ref{sec:FormationTimes}.}
\label{fig:MassZForm_Median}
\end{figure}
\begin{figure}
  \begin{center}
    \includegraphics[width=0.45\textwidth]{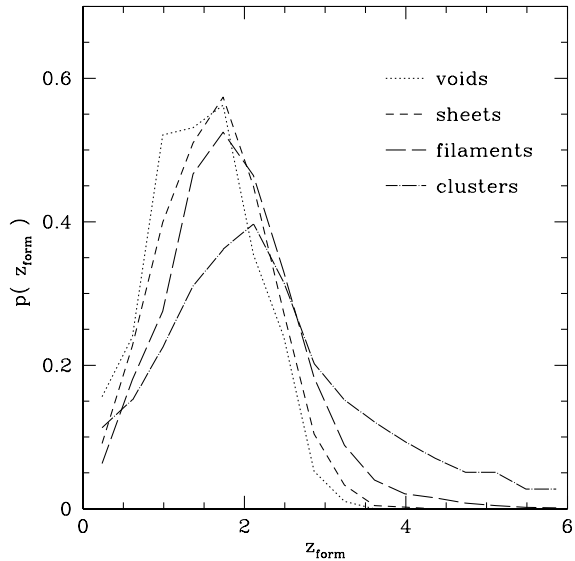}
  \end{center}
  \caption{The distribution of formation redshifts for haloes with masses $2\times10^{10}h^{-1}M_{\odot}<M<10^{11}h^{-1}M_{\odot}$ split into our four environment categories. Note that there are many more haloes in filaments than in clusters in this mass range.}
  \label{fig:ZFormPZForm}
\end{figure}

Both shape and scatter of the mass assembly curve depend strongly on the mass range at which they are evaluated.
This points to a strong relation between the formation redshift of haloes, their mass and environment. In Figure \ref{fig:MassZForm_Median} we plot the median formation redshift $z_{\rm med}$ as a function of halo mass $M$ for the haloes from our three simulations. Errorbars are estimates of the error in the median computed as
\begin{equation}
\Delta z_{\rm med} = \frac{z_{0.84}-z_{0.16}}{\sqrt{N_{\rm h}}},
\end{equation}
where $z_{0.84}$ and $z_{0.16}$ denote the 84th and 16th percentile of the distribution of $z_{\rm form}$, corresponding to the $1\sigma$ spread if the underlying distribution were Gaussian, and $N_{\rm h}$ is the number of haloes used to sample the distribution. For four decades in mass, ranging from $10^{10}h^{-1}\,M_{\odot}$ to $10^{14}h^{-1}\,M_{\odot}$, we find a tight logarithmic relation between halo mass and formation redshift, reflecting the hierarchical structure formation paradigm. Results for all boxes agree very well when considering haloes of at least 300 particles. We fit a function of the form
\begin{equation}
  z_{\rm med} = c_1\,-\,c_2\,\log_{10}\frac{M}{10^{12}h^{-1}\,M_{\odot}}.
\label{eq:FitMassZForm}
\end{equation}
The parameters given by a robust fit to all haloes from the three simulations  are:
\begin{eqnarray*}
c_1 & = & 1.29\pm0.07, \\
c_2 & = & 0.312\pm0.006.
\end{eqnarray*}
For haloes with masses between $10^{10}h^{-1}\,M_{\odot}$ and $\approx10^{12}h^{-1}\,M_{\odot}$ we find that $z_{\rm med}$ strongly depends on environment. This dependence increases, the lower the mass of the haloes. Our results are in very good agreement with \cite{Sheth2004}, \cite{Gao2005}, \cite{Harker2006} and \cite{Reed06}. These authors found that haloes of given mass but different formation epochs show different clustering properties. In particular, they have shown that small-mass haloes with higher formation times cluster more strongly and are thus most likely associated to denser environments. For haloes with masses $M<5\times 10^{12}h^{-1}\,M_{\odot}$ we again fitted relation (\ref{eq:FitMassZForm}) separately for cluster, filament, sheet and void environments. The slope parameters $c_2$  are significantly different for the four environments. A robust fit to the data combined from all three simulations yields the fit parameters 
\begin{eqnarray*}
%\begin{equation}
\begin{array}{lll}
\left. \begin{array}{rcl}
c_1 & = & 1.42\pm0.39 \\
c_2 & = & 0.54\pm0.03
\end{array} \right\} & \hspace{-6pt}\textrm{clusters, }&\hspace{-8pt}M<5\times10^{12}h^{-1}M_\odot; \\
\left. \begin{array}{rcl}
c_1 & = & 1.30\pm0.04 \\
c_2 & = & 0.39\pm0.01
\end{array} \right\} & \hspace{-6pt}\textrm{filaments, }&\hspace{-8pt}M<5\times10^{12}h^{-1}M_\odot;\\
\left. \begin{array}{rcl}
c_1 & = & 1.21\pm0.11 \\
c_2 & = & 0.28\pm0.01
\end{array} \right\} & \hspace{-6pt}\textrm{sheets, }&\hspace{-8pt}M<5\times10^{12}h^{-1}M_\odot; \\
\left. \begin{array}{rcl}
c_1 & = & 1.36\pm0.48 \\
c_2 & = & 0.08\pm0.04
\end{array} \right\} & \hspace{-6pt}\textrm{voids, }&\hspace{-8pt}M<5\times10^{11}h^{-1}M_\odot.
\end{array}
%\end{equation}
\end{eqnarray*}
For $M>5\times10^{12}h^{-1}\,M_{\odot}$, we do not find any dependence on environment and the relation between $z_{\rm med}$ and halo mass is best fit by the relation for all haloes given above.

The differences between the environments become even more significant when considering mean values of $z_{\rm form}$ instead of the medians due to the skewness of the formation redshift distributions in each mass bin. To illustrate this, we plotted in figure \ref{fig:ZFormPZForm} the distribution of formation redshifts for haloes with masses $2\times10^{10}h^{-1}M_{\odot}<M<10^{11}h^{-1}M_{\odot}$ in the four environments. In very good agreement with \citet{wang06}, we find that the oldest haloes with $z_{\rm form}>3$ are relatively overrepresented in cluster environments. Thus, small mass haloes in the vicinity of clusters tend to be older, and there has to be some effect that prevents them from strong continuous accretion and major mergers in this environment. In absolute numbers, we find a comparable amount of these very old low mass haloes also in our filament environments, such that, to a lesser extent, a similar effect must be present in filaments. \citet{wang06} suggest that the survival of these fossil haloes may be related to the ``temperature'' of the surrounding flow.  It is evident from these findings, that this ``temperature'' would then strongly correlate with the dimension of the stable manifold in our classification of environment. The precise connection has to be investigated in future work, but it is conceivable that the higher the number of stable dimensions, the less coherent and more accelerated is the infall of surrounding matter, and the stronger the heating of dark matter random motion.
\subsection{Halo Spin}
\label{sec:HaloSpins}
\begin{figure*}
  \begin{center}
    \includegraphics[width=0.45\textwidth]{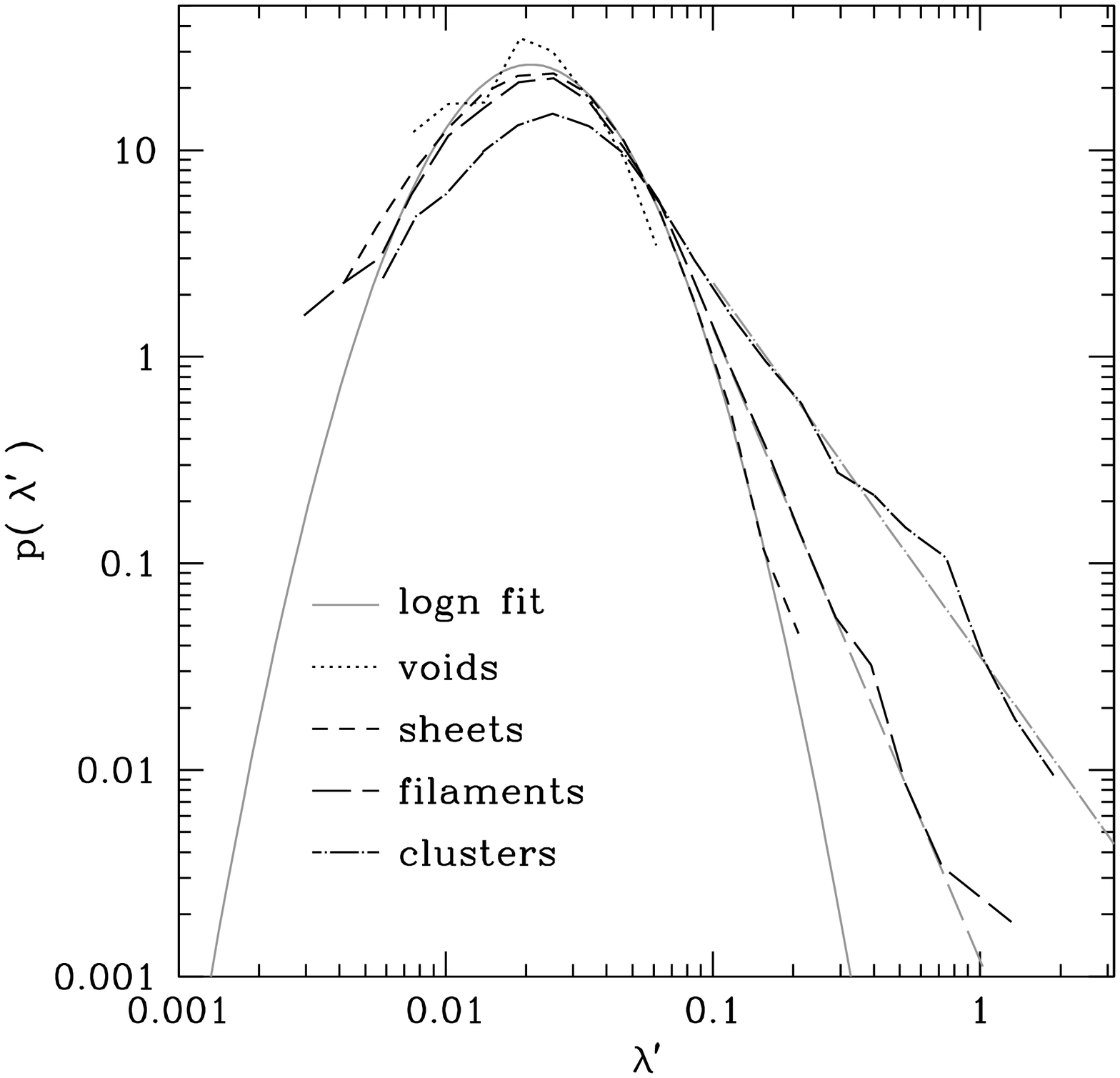}
    \includegraphics[width=0.45\textwidth]{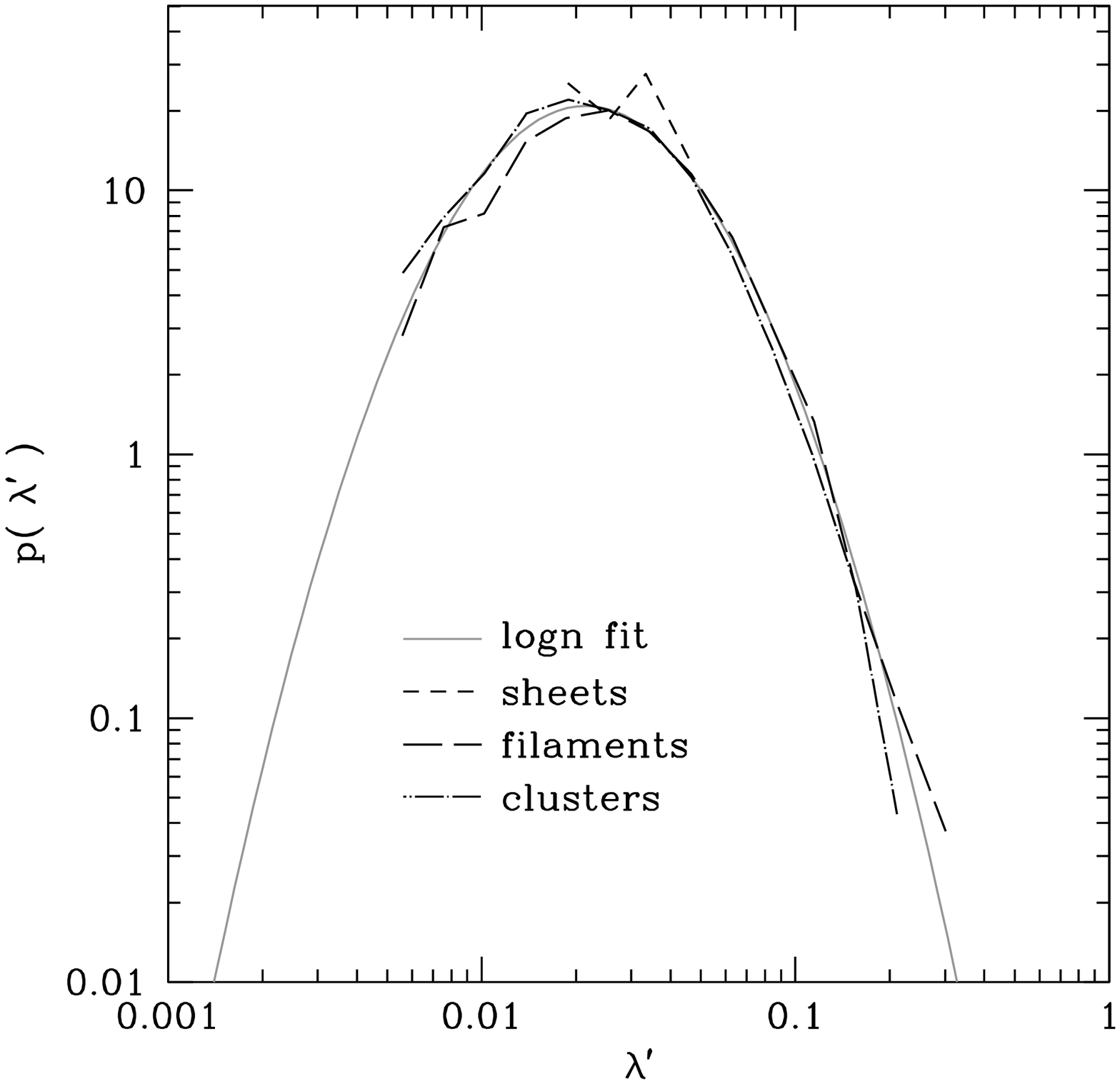}
  \end{center}
\caption{{\it Left panel}: Distribution of halo spin parameter $\lambda^\prime$ for haloes in the mass interval $5\times10^{10}h^{-1}M_{\odot} < M < 5\times10^{11}h^{-1}M_{\odot}$ residing in clusters, filaments, sheets and voids. Statistics are combined for all three simulation volumes. The solid grey line indicates the fit of a log-normal distribution to the sample $\lambda^{\prime}<0.1$, not split by environment. The dashed grey line shows a power-law fit to the distribution of $\lambda^{\prime}>0.1$ for haloes in filaments, the dash-dotted grey line the corresponding fit to cluster haloes. {\it Right panel}: Spin parameter distribution for halo masses $M>5\times10^{12}h^{-1}M_{\odot}$ for which haloes in voids are not present. The solid grey line shows the fit of a log-normal distribution to the whole sample, not divided into environment types. All fit parameters are given in section \ref{sec:HaloSpins}.}
  \label{fig:LambdaEnvMass}
\end{figure*}
We investigate the dependence of the halo spin parameter $\lambda^\prime$ on environment.
Figure \ref{fig:LambdaEnvMass} shows the distribution of $\lambda^\prime$ in the mass ranges $5\times10^{10}h^{-1}M_{\odot} < M < 5\times10^{11}h^{-1}M_{\odot}$ and $M>5\times 10^{12}h^{-1}M_{\odot}$.
In the high-mass bin, the distribution of spin parameters is well approximated by a log-normal probability density function,
\begin{equation}
p(\lambda^\prime) = \frac{1}{\lambda^\prime\,\sigma_{\lambda^\prime}\,\sqrt{2\pi}}\,\exp\left[ -\frac{\log^2\left( \lambda^\prime\,/\,\lambda^\prime_0 \right)}{2\,\sigma^2_{\lambda^\prime}}\right]\;,
\end{equation}
with best-fitting parameters  $\lambda^\prime_0=0.035$ and width $\sigma_{\lambda^{\prime}}=0.70$.
However, for $M<5\times 10^{12}h^{-1}M_{\odot}$, we find a tail of rapidly spinning haloes that is most prominent in clusters, and to a lesser extent in filaments. For the mass range $5\times10^{10}h^{-1}M_{\odot} < M < 5\times10^{11}h^{-1}M_{\odot}$, we find good agreement of all environments with a log-normal distribution only for spin parameters $\lambda^{\prime}<0.1$. The fit parameters for $\lambda^{\prime}<0.1$ in the low mass regime are $\lambda^\prime_0=0.030$ and $\sigma_{\lambda^{\prime}}=0.61$.  Our findings for the parameter $\lambda^{\prime}_0$ agree well with earlier findings \citep[e.g.][]{Bullock2001,Bett06}. At $\lambda^{\prime}\approx0.1$, however, we detect evidence for a departure from the log-normal distribution that is very well fit by a power-law behaviour. We find
\begin{equation}
p(\lambda^{\prime}\,|\,\lambda^{\prime}>0.1) = 0.0012\,\lambda^{\prime\,-3.1}
\end{equation}
for haloes in filaments and
\begin{equation}
p(\lambda^{\prime}\,|\,\lambda^{\prime}>0.1) = 0.035\,\lambda^{\prime\,-1.8}
\end{equation}
for haloes in clusters.
This tail is almost independent of the assumed value of $\alpha$, i.e. the fraction of the virial radius within which $\lambda^{\prime}$ is determined. However, the environmental dependence of the spin distribution slightly decreases when only the very innermost parts of a halo are used to determine $\lambda^\prime$. We have also verified that the high-spin tail of the distribution is not affected by measurement errors of the halo spin, i.e. the statistics remains unaltered when only haloes containing $>1000$ particles are considered. 

Our results appear to be in disagreement with \cite{AvilaReese2005} who found that haloes in clusters are less rapidly spinning than in the field.
However, a direct comparison is problematic since {\it a)} we use a different halo finder algorithm, {\it b)} we do not consider sub-haloes (which likely suffer strong tidal stripping), and {\it c)} we use a different definition of the cluster environment. However, we agree well with their finding that the parameter $\sigma_{\lambda^{\prime}}$ of the log-normal fit is significantly larger for haloes in cluster environments than for haloes in under-dense regions.
\begin{figure*}
\begin{center}
\includegraphics[width=0.45\textwidth]{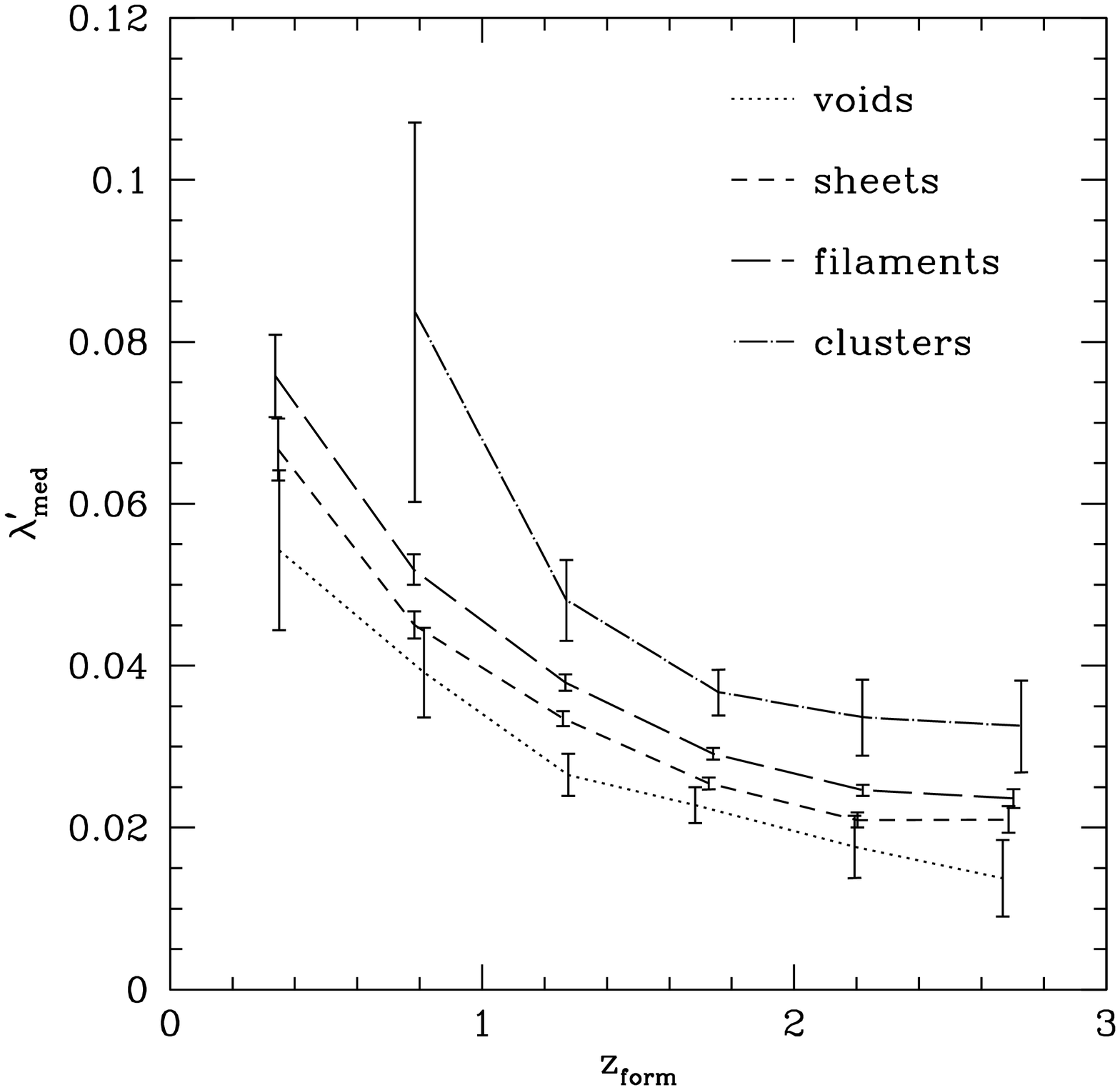}
\includegraphics[width=0.45\textwidth]{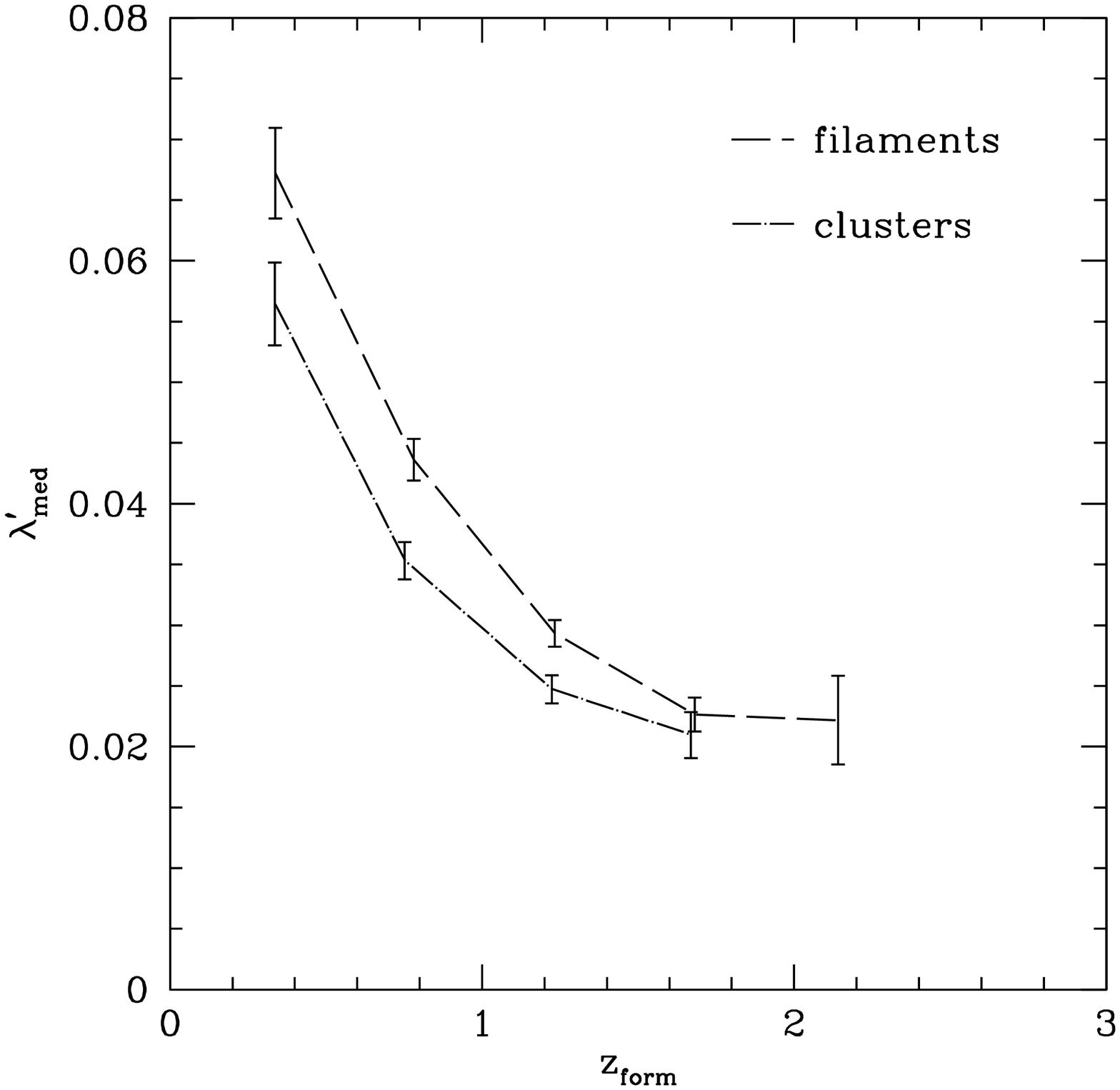}
\end{center}
\caption{The median spin parameter $\lambda^\prime$ of haloes in the mass range $5\times10^{10}h^{-1}M_{\odot} < M < 5\times10^{11}h^{-1}M_{\odot}$ (left) and $M > 5\times10^{12}h^{-1}M_{\odot}$ (right) for the four different environments as a function of their formation redshift. Errorbars indicate the $1\sigma$ uncertainty in the median.}
\label{fig:LambdaZForm}
\end{figure*}
\cite{Hetznecker06} have recently shown that the halo spin parameter increases significantly after a major merger, and relaxes to more standard values after 1-2 Gyr.
The formation redshift of a halo, as defined in section \ref{sec:Nbody}, is a good indicator for the occurrence of major mergers. Low formation redshifts correspond to recent major mergers, while high values of $z_{\rm form}$ denote less violent accretion histories. Figure \ref{fig:LambdaZForm} shows the median spin parameter as
a function of $z_{\rm form}$ for two mass-bins.  
We find that, in all environments and mass ranges, the median $\lambda^\prime$ is a decreasing function of $z_{\rm form}$. At the same time, for a given $z_{\rm form}$, the spin parameter shows an important environmental dependence: low-mass haloes tend to spin faster if they reside in clusters while massive haloes tend to spin slower in this environment.
The haloes with the largest spin parameter (median $\lambda^\prime>0.1$) are low mass haloes $M<5\times10^{12}h^{-1}M_\odot$ that reside in clusters and have $z_{\rm form}<1$. However, for fixed $z_{\rm form}$ haloes in clusters have higher median $\lambda^\prime$ compared to the other environments. 
\subsection{Angular Momentum Alignments}
\begin{figure}
  \begin{center}
    \includegraphics[width=0.45\textwidth]{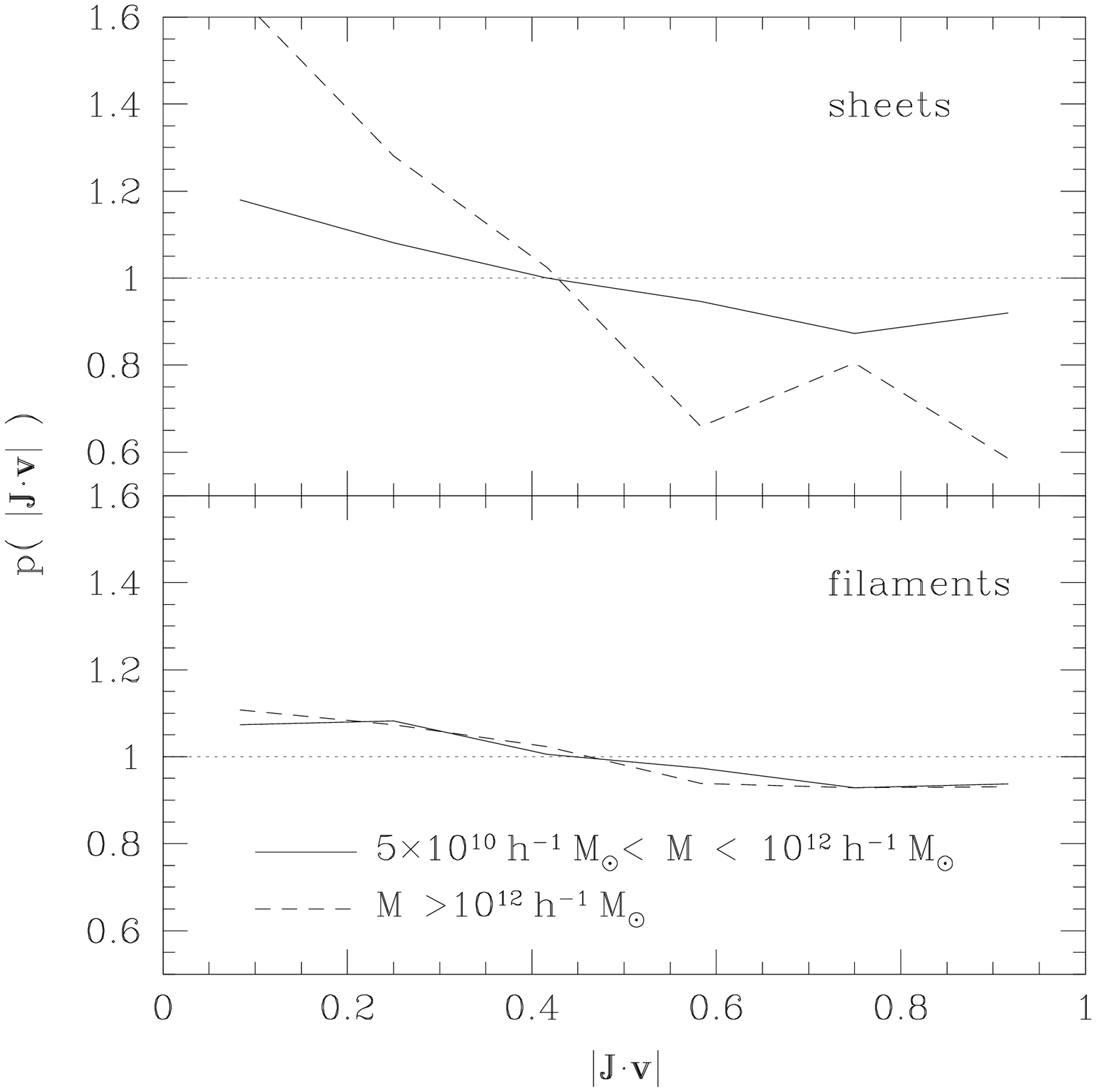}
  \end{center}
  \caption{The alignment between halo angular momentum vectors and the eigenvector corresponding a direction perpendicular to the sheets (top), and corresponding the direction of the filaments (bottom), for haloes in these two environments. Halo populations are divided in two bins  $5\times10^{10}h^{-1}M_{\odot}<M<10^{12}h^{-1}M_{\odot}$ and $M>10^{12}h^{-1}M_\odot$. The dotted grey lines indicate a random signal.}
  \label{fig:AM-Filament-Alignment}
\end{figure}
Do halo spin directions retain memory of the cosmic web in which the haloes formed?
Both filaments and sheets have a preferred direction given by the structure of the eigenspace. While filaments are one-dimensional structures with a preferred direction in space, sheets are two-dimensional and can thus be uniquely described by their normal vectors. Using the definition in section \ref{sec:ClassBasic} these directions are given by the unit eigenvector $\hat{\mathbf{v}}$ corresponding to the negative eigenvalue of the tidal field tensor for filaments and the positive eigenvalue for sheets.
One can therefore compute the degree of alignment between the angular momentum vector of a halo and the respective eigenvectors of the environment in which it resides, ${\hat{\mathbf{J}} \cdot \hat{\mathbf{v}}}$.
Figure \ref{fig:AM-Filament-Alignment} shows the distribution of alignments between halo angular momentum and both filament direction and sheet normal vector in the two mass-bins $5\times10^{10}h^{-1}M_{\odot}<M<10^{12}h^{-1}M_{\odot}$ and $M>10^{12}h^{-1}M_{\odot}$. For haloes in filaments we find only a weak trend for their angular momenta to be aligned with the filament direction. Haloes in sheets, however, show a very strong tendency to have their angular momentum parallel to the sheet. Similar correlations
are also found in walls delimiting voids \citep{Patiri06,Brunino06}, 
and might be reflected in the distribution of galactic disks \citep[e.g.][]{Navarro04,Trujillo06}.
We did not detect any strong correlation with eigenvectors of the other environments. The presence of alignments between large-scale structures and halo spins could produce a coherent alignment of galaxy shapes and thus generate a systematic contamination in weak lensing maps of cosmic shear \citep[e.g.][]{Hirata04, Heymans06}.

We next compute correlations of the intrinsic angular momentum of each halo with both the intrinsic angular momentum and the orbital angular momentum of neighbouring haloes residing in the same environment. We define the spin-spin correlation function as \citep{Porciani02,Bailin2005}:
\begin{equation}
\xi_{\mathbf{J}\cdot\mathbf{J}}(r) = \langle \hat{\mathbf{J}}({\bf x})\cdot\hat{\mathbf{J}}({\bf x}+{\bf r})\rangle,
\end{equation}
where $\mathbf{J}$ is the intrinsic angular momentum of each halo, and the average is taken over all pairs
of haloes which are separated by a distance $r$ and reside in the same environment class. Similarly, we define the spin-orbit correlation as
\begin{equation}
\xi_{\mathbf{J}\cdot\mathbf{L}}(r) = \langle \hat{\mathbf{J}}({\bf x})\cdot\hat{\mathbf{L}}({\bf x}+{\bf r})\rangle,
\end{equation}
where $\mathbf{L}$ is the relative orbital angular momentum between two haloes separated by a distance $r$.
\begin{figure}
  \begin{center}
    \includegraphics[width=0.45\textwidth]{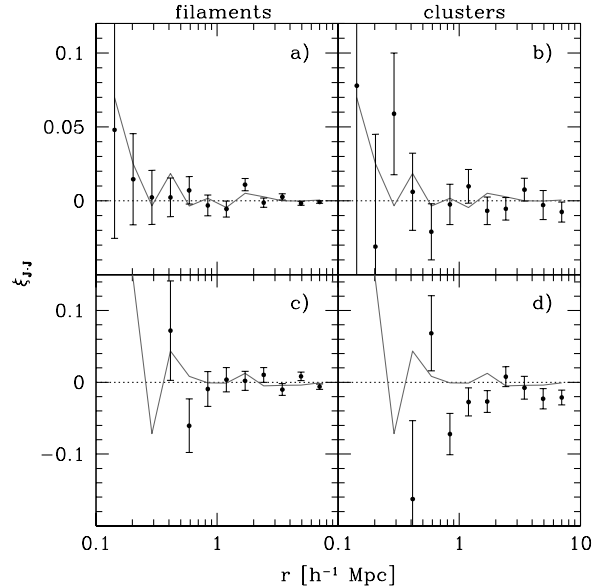}
  \end{center}
  \caption{The mean alignment of intrinsic spin angular momenta between haloes in filaments, panels a) and c), and clusters, panels b) and d). Data is plotted for the two mass-bins $5\times10^{10}h^{-1}M_{\odot} < M < 5\times10^{11}h^{-1}M_{\odot}$, panels a) and b), and $M > 5\times10^{12}h^{-1}M_{\odot}$, panels c) and d). Errorbars are the $1\sigma$ uncertainty of the mean. The grey line indicates the mean correlation for the whole halo population, independent of environment, for the same mass bins. The dotted line represents a random signal with no correlation.}
\label{fig:AMAM_JJ}
\end{figure}
\begin{figure}
  \begin{center}
    \includegraphics[width=0.45\textwidth]{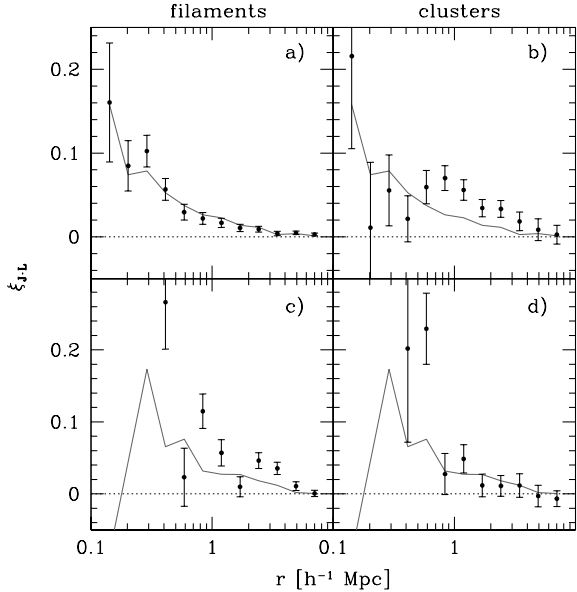}
  \end{center}
  \caption{The mean alignment of intrinsic spin and relative orbital angular momentum between haloes in filaments, panels a) and c), and clusters, panels b) and d). Data is plotted for the two mass-bins $5\times10^{10}h^{-1}M_{\odot} < M < 5\times10^{11}h^{-1}M_{\odot}$, panels a) and b), and $M > 5\times10^{12}h^{-1}M_{\odot}$, panels c) and d). Errorbars are the $1\sigma$ uncertainties of the mean. The solid grey line indicates the mean correlation for the whole halo population, independent of environment, for the same mass bins. The dotted line represents a random signal with no correlation.}
\label{fig:AMAM_JL}
\end{figure}
Figure \ref{fig:AMAM_JJ} shows the spin-spin correlation for haloes in two mass bins, $5\times10^{10}h^{-1}M_{\odot} < M < 5\times10^{11}h^{-1}M_{\odot}$ and $M > 5\times10^{12}h^{-1}M_{\odot}$. We find a significant correlation only for haloes with $M>5\times10^{12}h^{-1}M_\odot$ in cluster environments. These haloes have a strong tendency to have their spin vectors anti-parallel to the spins of haloes within a distance of a few Mpc. All other correlations are essentially consistent with a random signal. The results for haloes with masses $5\times10^{11}h^{-1}M_{\odot} < M < 5\times10^{12}h^{-1}M_{\odot}$ are fully consistent with those for the lower mass bin and therefore not shown in the plots.
Regarding the alignment of spin and orbital angular momenta, the results, given in Figure \ref{fig:AMAM_JL}, 
show a much stronger signal and a clear dependence on environment. We find an evident tendency for the two angular momenta to be {\it parallel} regardless of mass and environment. Remarkably this correlation significantly extends out to $\sim 2\,h^{-1}$ Mpc in all environments and is most prominent for smaller haloes in clusters and massive haloes in filaments. 

%%%

\section{Summary}
\label{conclusions}
We have presented a new method to classify dark-matter haloes as belonging to four different environments:
{\it clusters}, {\it filaments}, {\it sheets} and {\it voids}.
This scheme computes the dimensionality of the stable manifold for the orbits of test particles
by simply looking at the number of positive eigenvalues of the local tidal tensor.
The algorithm contains only one free parameter: the smoothing radius for the gravitational potential.
This quantity fixes
the length-scale over which the stability of structures is determined and 
can be fine tuned to optimise the classification. 
At the same time, combining the results obtained adopting two or more different smoothing scales 
allows us to select regions
with particular properties in the large-scale structure (e.g. transition regions between the basic four environments).

Our classification scheme correlates with local density so that the densest regions are always 
associated with clusters and the emptiest with voids.
However, our method retains more information on the local dynamics  
and a simple halo classification based on density will unavoidably mix our populations up.

We have used the classification scheme to study how the properties of isolated
dark-matter haloes depend on the environment
in which they reside at $z=0$. Our main results can be summarised as follows.

\vspace{2mm}
\noindent{\it 1) Halo shapes}
\begin{itemize}
\item Massive haloes with $M>\,{\rm a\ few}\,\times 10^{12} h^{-1} M_\odot$ do not show any
significant dependence of their shape on environment.
\item Less massive haloes in 
clusters are less spherical and more oblate than in other regions
but the trend is generally weak compared with the intrinsic scatter. 
\end{itemize}

\noindent{\it 2) Halo formation times}
\begin{itemize}
\item
For the whole halo population (not split by environment) we found a very strong 
correlation between median formation redshift and halo mass. A fit to this relation 
which holds for halo masses between $10^{10} h^{-1} M_\odot<M<10^{14} h^{-1} M_\odot$
is given in equation (\ref{eq:FitMassZForm}).
This dependence is a direct consequence of hierarchical structure formation. 
\item
For $M<5\times10^{12}h^{-1}M_\odot$ haloes of fixed mass
in the four environments have significantly different mass assembly histories. 
In particular, cluster haloes tend to be older while void haloes younger.
All this hints at mechanisms that suppress the growth of lower 
mass haloes in clusters and lead to an enhanced survival rate of fossil haloes
\citep[see e.g.][for a possible explanation]{wang06}.
\item
Analytic fitting formulae for the dependence of the median formation redshift on halo mass and environment
are given in Section \ref{sec:FormationTimes}.
\end{itemize}

\noindent{\it 3) Halo spins}
\begin{itemize}
\item
The median spin parameter of all haloes is the highest in clusters 
followed in order by filaments, sheets and voids.  
This dependence, presumably, has its origin in the tidal-torque history of the haloes
which likely correlates with the  
specific eigenstructure of the tidal field at the final halo position
\citep{Bond96, Porciani02, Porciani02b}. 
\item
This trend is reversed for massive objects. 
Haloes with $M>5\times10^{12}h^{-1}M_\odot$ in clusters are less 
rapidly spinning than in filaments.
\item 
On the other hand, for smaller masses,
haloes in clusters generally possess higher spin parameters than in the 
other three environments. 
As these rapidly spinning haloes have also the most recent formation time, we 
conjecture that the high spin tail is generated by recent major mergers that bias the distribution towards 
rapid rotation. Hence, the high-spin tail of unrelaxed haloes overlaps the distribution of quiescently evolving 
haloes which is best fit by a log-normal distribution. 
\end{itemize}

\noindent{\it 4) Alignment of halo spins and large-scale structures} 
\begin{itemize}
\item
Haloes in sheets show a strong tendency for their spin vector to lie in the symmetry plane
of the mass distribution. 
This effect is present for all haloes but it becomes much more prominent for haloes with  
$M<10^{12}h^{-1}M_\odot$. 
\item 
For haloes in filaments, there is a slightly enhanced probability to find their angular momentum orthogonal 
to the filament direction, independently of mass.
\item 
No other significant correlation has been detected (but we suffer from small-number statistics in voids).
\end{itemize}

\noindent{\it 5) Spatial correlations between halo angular momenta}
\begin{itemize}
\item
Significant spin-spin correlations have been only detected for massive haloes in clusters.
In this case, haloes in close pairs (separations smaller than a few Mpc) show a weak
tendency to have antiparallel spins.
\item
Alignments between spin and orbital angular momentum, however, were found to
be much stronger. Regardless of mass and environment, spins of haloes in close pairs 
tend to be preferentially parallel to the orbital angular momentum of the pair.
This strong effect is even enhanced for low-mass haloes in clusters and massive haloes in filaments.
\end{itemize}

Our study has revealed that a number of halo properties depend on environment.
This shows that our dynamical classification is physical  
and represents a first step towards understanding how the galaxy formation process
is influenced by large-scale structures. 
We will further explore the potential of this method in future work.

\section*{Acknowledgements}
OH acknowledges support from the Swiss National Science Foundation.
All simulations were performed on the Gonzales cluster at ETH Zurich, Switzerland.
We are very grateful to Olivier Byrde for excellent cluster support.
While our paper was under consideration for publication, 
an interesting related work by \cite{Aragon06} appeared as a preprint.

%\bibliography{HPC06}

\label{lastpage}

\end{document}